\documentclass[11pt,a4paper]{article}
\usepackage{amsmath,amssymb}
\usepackage{epsfig,graphicx}

\begin{document}

\title{\bf Inclusive jet production at Tevatron
in the Regge limit of QCD}
\author{V.~A.~Saleev$^{a,b}$\footnote{{\bf e-mail}: saleev@ssu.samara.ru},
A.~V.~Shipilova$^{a}$\footnote{{\bf e-mail}:
alexshipilova@ssu.samara.ru}, E.~V.~Yatsenko$^{a}$\footnote{{\bf
e-mail}: spectrum11@mail.ru}
\\
$^a$ \small{\em Samara State University} \\
\small{\em Ac. Pavlova, 1, Samara 443011, Russia}\\
$^b$ \small{\em Samara State Aerospace University} \\
\small{\em Moscow Highway, 34, Samara 443086, Russia }}
\date{}
\maketitle

\begin{abstract}
We consider inclusive hadroproduction of jets,  prompt photons,
$b-$jets, and $D$-mesons in the quasi-multi-Regge-kinematics
approach based on the hypothesis of gluon and quark Reggeization in
$t$-channel exchanges at high energies. The data taken by the CDF
and D0 Collaborations at the Fermilab Tevatron are well described
without adjusting parameters. We find the main contribution to
inclusive jet production to be the scattering of two Reggeized
gluons, described by the effective Reggeon-Reggeon-gluon vertex, and
to $b$-jet and $D$-meson production --- the scattering of a
Reggeized gluon and a Reggeized quark to an ordinary quark, being
expressed in terms of the Reggeon-Reggeon-quark vertex. The main
contribution to prompt photon production arises from Reggeized
quark-Reggeized antiquark annihilation, which is described by the
effective Reggeon-Reggeon-photon vertex. Our analysis is based on
the Kimber-Martin-Ryskin prescription for unintegrated gluon and
quark distribution functions using as input the
Martin-Roberts-Stirling-Thorne collinear parton distribution
functions of the proton.
\end{abstract}

\section{Introduction}
The study of inclusive jet and prompt photon production with large
transverse momenta coming from a hard interaction between two
partons in hadron collisions at high-energy colliders is of great
interest for the test of perturbative quantum chromodynamics
(QCD). Also these processes provide an opportunity to extract
information on the parton densities within protons.

The total center-of-mass (CM) energy at the Tevatron Collider,
$\sqrt{S}=1.8$~TeV in Run~I and $\sqrt{S}=1.96$~TeV in Run~II,
sufficiently exceeds the scale $\mu$ of the relevant hard
processes, so that $\sqrt{S}\gg \mu \gg \Lambda_{\rm QCD}$, where
$\Lambda_{\rm QCD}$ is the asymptotic scale parameter of QCD. In
such a high-energy regime, the contributions from subprocesses
involving $t$-channel exchanges of partons (gluons and quarks) to
the production cross section may become dominant. Then, the
transverse momenta of the incoming partons and their off-shell
properties can no longer be neglected, and we deal with {\it
Reggeized} $t$-channel partons. In this so-called
quasi-multi-Regge kinematics (QMRK), the particles (multi-Regge)
or groups of particles (quasi-multi-Regge) produced in the
collision are strongly separated in rapidity space. In the case of
inclusive jet and prompt photon production, this implies the
following: a single jet or photon is produced in the central
region of rapidity, while other particles are produced at large
rapidities.  The QMRK approach \cite{QMRK} is particularly
appropriate for this kind of high-energy phenomenology. It is
based on an effective quantum field theory implemented with the
non-Abelian gauge-invariant action including fields of Reggeized
gluons \cite{Lipatov95} and quarks \cite{LipatoVyazovsky}. Roughly
speaking, the Reggeization of amplitudes is a trick that offers an
opportunity to take into account efficiently  large radiation
corrections to the processes under Regge limit condition beyond
the collinear approximation of the parton model. The particle
Reggeization is known in high-energy quantum electrodynamics (QED)
for electrons only \cite{GellMann} and for gluons and quarks in
QCD \cite{BFKL,QR1old,QR2old}.

In this paper, we consider inclusive hadroproduction of jets,
$b-$jets, $D$-mesons and prompt photons in the QMRK approach
applying the hypothesis of gluon and quark Reggeization
\cite{BFKL,QR1old,QR2old}.

\section{Effective vertices and amplitudes of $2\to 1$ subprocesses}
Throughout our analysis we will use the following definitions:
$\mathcal{Q}$
--- Reggeized quark, $\mathcal{R}$
--- Reggeized gluon, $q$ --- ordinary quark, $g$ --- Yang-Mills
gluon and indicate the 4-momenta of the particles in parenthesis
after their denotations.

We consider high transverse momenta particle production in
proton-antiproton collisions at the Tevatron Collider. In the CM
frame the 4-momenta of the incoming proton and antiproton are
written down as $P_1=E_1(1,0,0,1)$ and $P_2=E_2(1,0,0,-1)$,
correspondingly.

We also introduce the ancillary vectors $n_\mu^\pm=(1,0,0,\mp 1)$
and for any 4-momentum $k^\mu, q^\mu, p^\mu$ define:
\begin{equation}
k^{\pm}=k\cdot
n^\pm,\quad\gamma^{(\pm)}_\mu(q,p)=\gamma_\mu+\not{q}\frac{n^{\pm}_\mu}{p^{\pm}},
\end{equation}
\begin{equation}
\gamma_\mu^{(+-)}(q_1,q_2)=\gamma_\mu
-\frac{\not{q}_1n^{-}_{\mu}}{q_2^-}
-\frac{\not{q}_2n^{+}_{\mu}}{q_1^+}.
\end{equation}

In the CM frame the 4-momenta of a Reggeized particle coming from
initial hadron may be represented in the form:
$q_i=x_iP_i+q_{iT}$, $i=1,2$, where $x_i$
--- the longitudinal momentum fraction of initial hadron carried by the Reggeized particle,
$q_{iT}=(0,\mathbf{q_{iT}},0)$ --- the transverse momentum of the
latter. To make subsequent formulas shorter we define
$t_i=-q_{iT}^2=\mathbf{q_{iT}}^2$.

Below we present effective vertices and squared amplitudes of the
$2\to 1$ subprocesses with two Reggeized partons in the initial
state, they read: $\mathcal{Q}+\mathcal{R}\to q$,
$\mathcal{R+R}\to g$, $\mathcal{Q+\bar Q}\to g$, $\mathcal{Q+\bar
Q}\to \gamma$. Thus, the transverse momenta of the particles
involved in these subprocesses are related by the condition:
${\mathbf k}_T^2=t_1+t_2+2\sqrt{t_1t_2}\cos \phi_{12}$, where
$k=(k_0,\mathbf{k_T},k_z)$ is a 4-momentum of the final quark or
gluon,
$\phi_{12}$ --- the azimuthal angle between the vectors ${\mathbf
q}_{1T}$ and ${\mathbf q}_{2T}$.

At first, we consider hard subprocess of a single quark production
via Reggeized gluon-Reggeized quark scattering:
\begin{equation}
\mathcal{Q}(q_1)+\mathcal{R}(q_2)\to q(k), \label{sp:QRq}
\end{equation}
As the subject of our investigation is a high transverse momenta
jet production, we are allowed to neglect quark masses, so that
the effective vertex of the subprocess (\ref{sp:QRq}) may be
displayed like following, according to \cite{LipatoVyazovsky}:
\begin{equation}
C_{\mathcal{QR}}^q(q_1,q_2)=i\sqrt{4\pi\alpha_s}\,T^a\bar
U(k)\gamma^{(-)\mu}(q_1,q_2)\Pi_\mu^{(+)}(q_2)\label{amp:QRq},
\end{equation}
where $\alpha_s$ is the strong coupling constant, $T^a$ --- the
color gauge group SU(3) generators, $a=1,\ldots,N_c^2-1$,
$\displaystyle
 \Pi^{(+)}_\mu(q_2)=-\frac{q_2^-
n^+_\mu}{2\sqrt{t_2}}$. The corresponding squared amplitude has a
simple form \cite{Saleev2008}:
\begin{eqnarray}
\overline{|{\cal M}(\mathcal{Q}+\mathcal{R}\to
q)|^2}=\frac{2}{3}\pi \alpha_s {\mathbf k}_T^2. \label{sqamp:QRq}
\end{eqnarray}

The effective vertex of a transition of two Reggeized gluons to
Yang-Mills gluon $\mathcal{R+R}\to g$ may be written as
\cite{QMRK,BFKL,RRgold}:
\begin{eqnarray}
C_{\mathcal{RR}}^{g,\mu}(q_1,q_2)&=&-\sqrt{4\pi\alpha_s}f^{abc}\frac{q_1^+q_2^-}{2\sqrt{t_1t_2}}
\biggl[\left(q_1-q_2\right)^\mu+\nonumber\\&+&\frac{(n^+)^\mu}{q_1^+}\left(q_2^2+q_1^+q_2^-
\right)-
\frac{(n^-)^\mu}{q_2^-}\left(q_1^2+q_1^+q_2^-\right)\biggr],\label{amp:RRg}
\end{eqnarray}
where $a$ and $b$ --- color indices of Reggeized gluons with
4-momenta $q_1$ and $q_2$, correspondingly. The squared amplitude
of the subprocess $\mathcal{R}+\mathcal{R}\to g$ can be obtained
by simple calculations:
\begin{eqnarray}
\overline{|{\cal M}(\mathcal{R}+\mathcal{R}\to
g)|^2}=\frac{3}{2}\pi \alpha_s {\mathbf k}_T^2. \label{sqamp:RRg}
\end{eqnarray}

The effective vertex of the transition of a Reggeized quark with
4-momenta $q_1$ and a Reggeized antiquark with 4-momenta $q_2$ to
photon or Yang-Mills gluon looks like following
\cite{QR1old,QR2old}:
\begin{eqnarray}
C_{\mathcal{Q\bar Q}}^{\gamma (g),\mu}(q_1,q_2)&=&{\cal C}_{\gamma
(g)} \biggl[\gamma^\mu-\not q_1\frac{(n^-)^\mu}{q^-_1+q^-_2}-\not
q_2\frac{(n^+)^\mu}{q^+_1+q^+_2}\biggr],\label{amp:QQg}
\end{eqnarray}
where ${\cal C}_\gamma=-i\sqrt{4\pi\alpha}Z_q$ for photon,
$\alpha\simeq 1/137$
--- the fine structure constant and $Z_q$ --- quark electric charge.
In the case of gluon ${\cal C}_g=-i\sqrt{4\pi\alpha_s}T^a$. The
squared amplitude of the concerned transition has the form:
\begin{equation}
\overline{|{\cal M}\bigl(\mathcal{Q}+\bar{\mathcal{Q}}\to
\gamma(g)\bigr)|^2}={\cal B}_{\gamma (g)} (t_1+t_2),
\label{sqamp:QQg}
\end{equation}
where the factor ${\cal B}_\gamma=\dfrac{4}{3}\pi\alpha Z_q^2$ for
photon and ${\cal B}_g=\dfrac{16}{3}\pi \alpha_s$ for gluon.

It is evident from cited above formulas, that squared amplitudes
of $2\to 1$ subprocesses are equal to zero in the collinear
approximation, in which case we need to start from $2\to 2$
subprocesses, being of the next order in  $\alpha_s$.

Exploiting the hypothesis of high-energy factorization, we may
write a hadronic cross sections $d\sigma$ as convolutions of
partonic cross sections $d\hat\sigma$ with unintegrated parton
distribution functions (PDFs) $\Phi_a^h$ of Reggeized partons $a$
in the hadrons $h$. The unintegrated PDFs $\Phi_a^h(x,t,\mu^2)$
are related to their collinear counterparts $F_a^h(x,\mu^2)$ by
the normalization condition
\begin{equation}
xF_a^h(x,\mu^2)=\int^{\mu^2}dt\,\Phi_a^h(x,t,\mu^2),
\end{equation}
which yields the correct transition from formulas in the QMRK
approach to those in the collinear parton model, where the
transverse momenta of the partons are neglected.
In our numerical analysis, we adopt
the prescription proposed by Kimber, Martin, and Ryskin \cite{KMR}
to obtain unintegrated gluon and quark distribution functions for
the proton from the conventional integrated ones, as implemented
in Watt's code \cite{Watt}. As input for this procedure, we use
the Martin-Roberts-Stirling-Thorne \cite{MRST} proton PDFs.

In our analysis the renormalization and factorization scales are
identified and chosen to be $\mu=\xi k_T$, where $\xi$ is varied
between 1/2 and 2 about its default value 1 to estimate the
theoretical uncertainty. The resulting errors are indicated as
shaded bands in the figures.

As we consider production of particles with high transverse
momenta $k_T \gg m_q$, it is allowed to work in massless
approximation.

\section{Inclusive jet production}
The first subject of our investigation is inclusive jet
production. It was experimentally studied by CDF \cite{RRgCDF} and
D0 \cite{RRgD0} Collaborations in several rapidity intervals at
jet transverse momenta up to 700 GeV. The main contribution to
such processes in leading order (LO) of QMRK comes from $2\to 1$
subprocess of Reggeized gluon fusion producing Yang-Mills gluon,
which is described by the effective vertex (\ref{amp:RRg}) and
squared amplitude (\ref{sqamp:RRg}). Using the hypothesis of
high-energy factorization, we have
\begin{eqnarray}
d\sigma(p\bar p\to jet X)=\int\frac{dx_1}{x_1}\int
\frac{d^2q_{1T}}{\pi}\int\frac{dx_2}{x_2}\int
\frac{d^2q_{2T}}{\pi}\Phi^p_{g}(x_1,t_1,\mu^2)\Phi^{\bar
p}_{g}(x_2,t_2,\mu^2) d\hat\sigma(\mathcal{RR}\to
g)\label{eq:XSRRg}
\end{eqnarray}
For the reader's convenience, we present here compact formula for
the differential cross section (\ref{eq:XSRRg}):
\begin{eqnarray}
\frac{d\sigma}{dk_T\,dy}(p\bar p\to jet X)=\frac{1}{k_T^3}\int
d\phi_1\int dt_1\Phi^p_g(x_1,t_1,\mu^2)\Phi^{\bar
p}_g(x_2,t_2,\mu^2)  \overline{|{\cal M}(\mathcal{RR}\to g)|^2},
\label{eq:cXSRRg}
\end{eqnarray}
where $y$ is the rapidity, $\phi_1$ is the azimuthal angle
enclosed between the vectors ${\mathbf q}_{1T}$ and ${\mathbf
k}_T$,
\begin{equation}
x_{1,2}=\frac{k_T\exp(\pm y)}{\sqrt{S}},\qquad
t_2=t_1+k_T^2-2k_T\sqrt{t_1}\cos\phi_1.
\end{equation}

\begin{figure}[ht]
\begin{center}
\includegraphics[width=.535\textwidth, clip=]{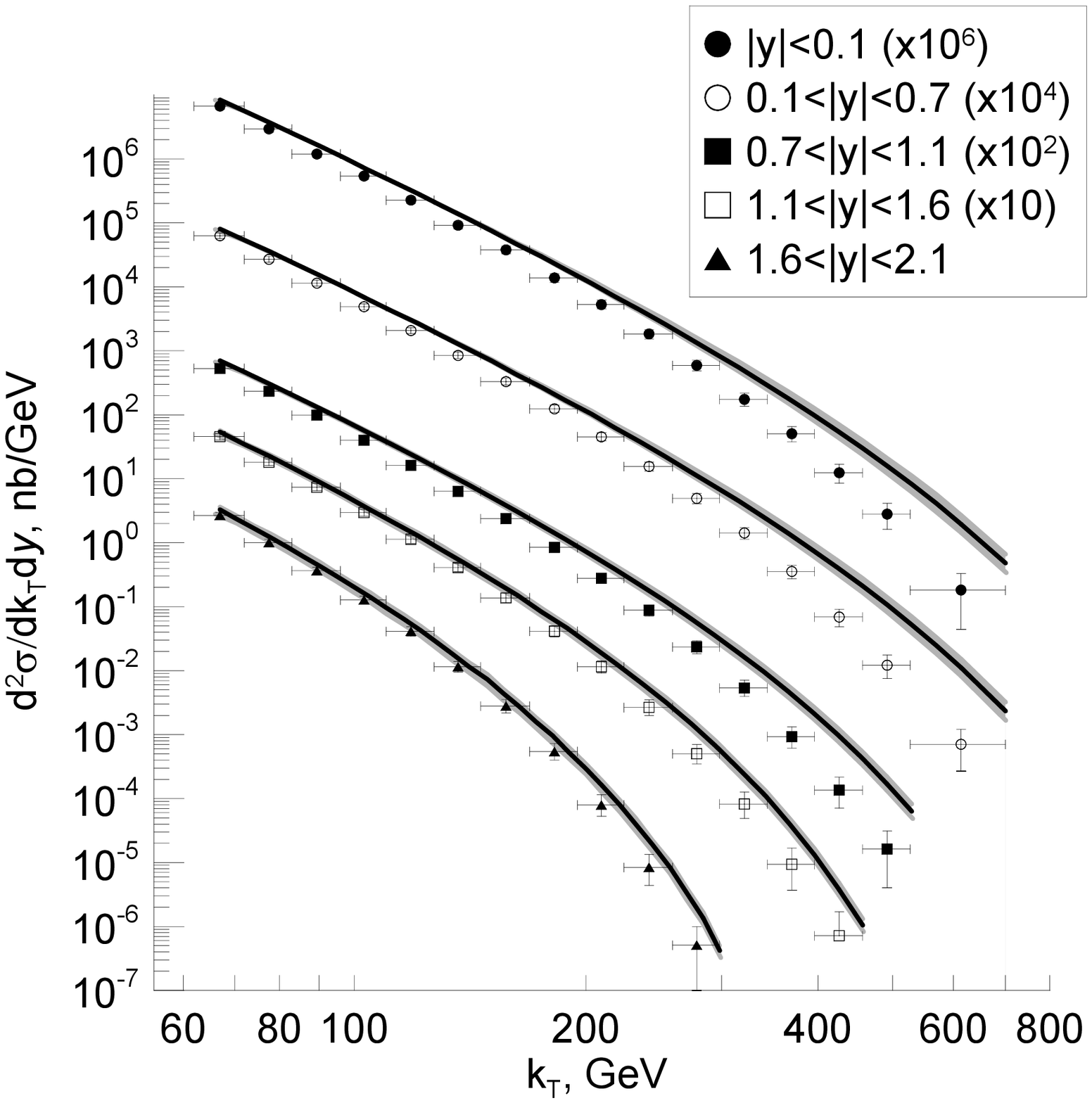}
\includegraphics[width=.535\textwidth, clip=]{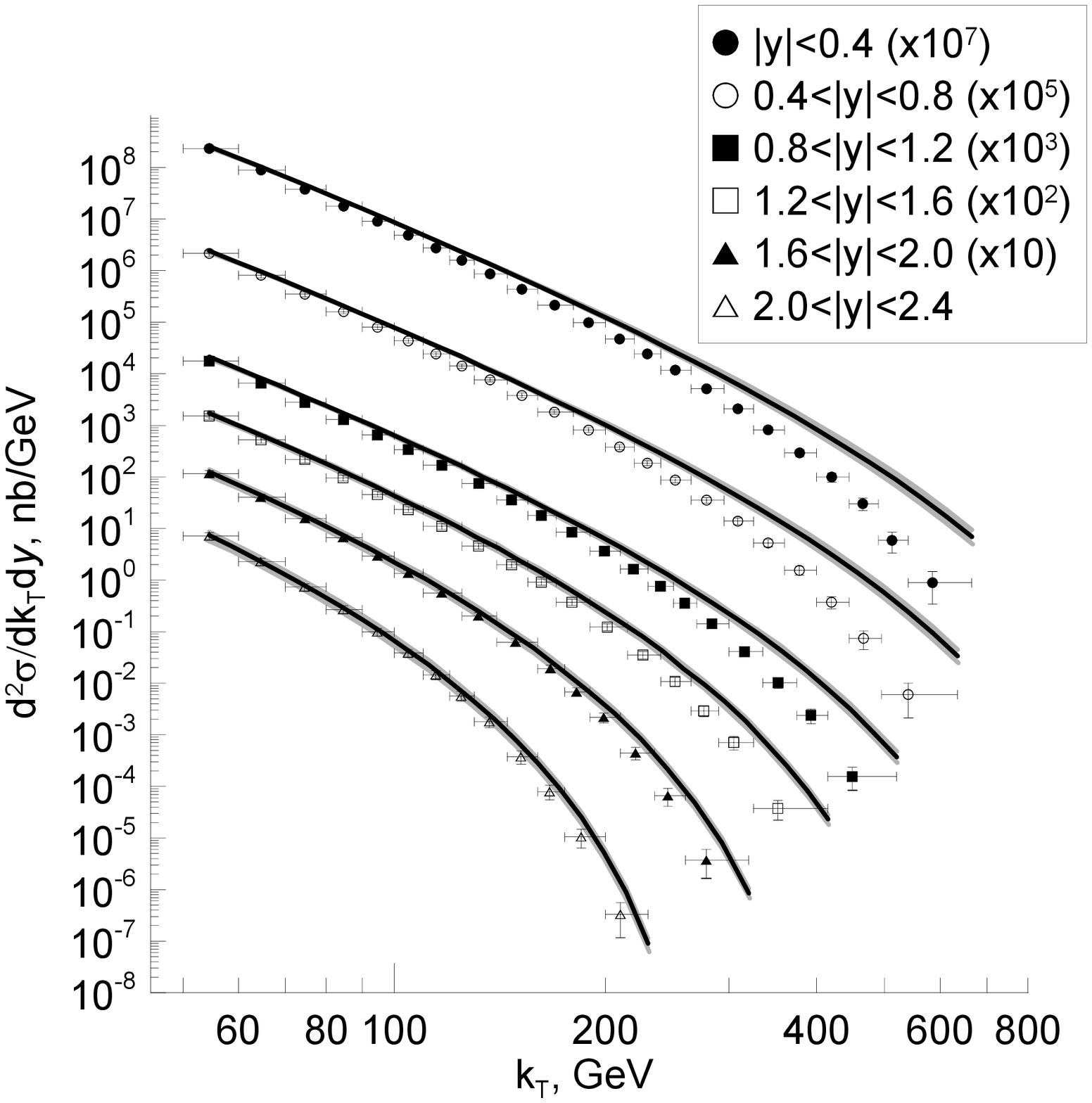}
\end{center}
\caption{\label{fig:1}%
QMRK predictions on inclusive jet production in comparison with
CDF \cite{RRgCDF} (top) and D0 \cite{RRgD0} (bottom) data.}
\end{figure}

In the Figs. \ref{fig:1}, top, and \ref{fig:1}, bottom, our
predictions obtained in the QMRK approach are compared with CDF
\cite{RRgCDF} and D0 \cite{RRgD0} data, correspondingly. The
theoretical predictions nicely agree with experimental data up to
200 GeV. The discrepancy at $k_T>200$~GeV arises because the
average values of the scaling variables $x_1$ and $x_2$ in the
unintegrated PDFs exceed 0.2, so that, strictly speaking, the QMRK
approach ceases to be valid.

\section{Inclusive $b$-jet production}
Recently CDF Collaboration presented preliminary data on inclusive
single $b$-jet production in $p\bar p$-collisions at Tevatron
Run~II \cite{CDF1}. The measurement was performed in the kinematic
range $38<k_T<400$~GeV and $|y|<0.7$. As the modulus of the
$b$-quark transverse momentum  $k_T\geq 32$~GeV \cite{CDF1},
sufficiently exceeds its mass $m_b$, it is justified to assume
beauty to be an active flavor in the proton. Such a way we are
allowed to examine the process under consideration in the
fixed-flavor-number scheme with $n_f=5$ active quark flavors. To
LO in the QMRK approach, there is only one partonic subprocess,
namely
\begin{equation}
\mathcal{Q}_b(q_1)+\mathcal{R}(q_2)\to b(k), \label{equ:RQ}
\end{equation}
where $\mathcal{Q}_b$ --- the Reggeized $b$ quark. Neglecting
Reggeized-quark masses, the effective vertex and squared amplitude
of the concerned transition are given by (\ref{amp:QRq}) and
(\ref{sqamp:QRq}). The high-energy factorization formula for the
cross section takes the form:
\begin{eqnarray}
d\sigma(p\bar p\to b X)&=&\int\frac{dx_1}{x_1}\int
\frac{d^2q_{1T}}{\pi}\int\frac{dx_2}{x_2}\int
\frac{d^2q_{2T}}{\pi} \left[\Phi^p_{b}(x_1,t_1,\mu^2)\Phi^{\bar
p}_{g}(x_2,t_2,\mu^2)+\right.
\nonumber\\
&&{}+\left.\Phi^p_{g}(x_1,t_1,\mu^2)\Phi^{\bar p}_{b}
(x_2,t_2,\mu^2)\right]d\hat\sigma(\mathcal{Q}_b\mathcal{R}\to b),
\end{eqnarray}
and it may be written compactly in the way similar to
(\ref{eq:cXSRRg}).

\begin{figure}[ht]
\begin{center}
\includegraphics[width=.6\textwidth, clip=]{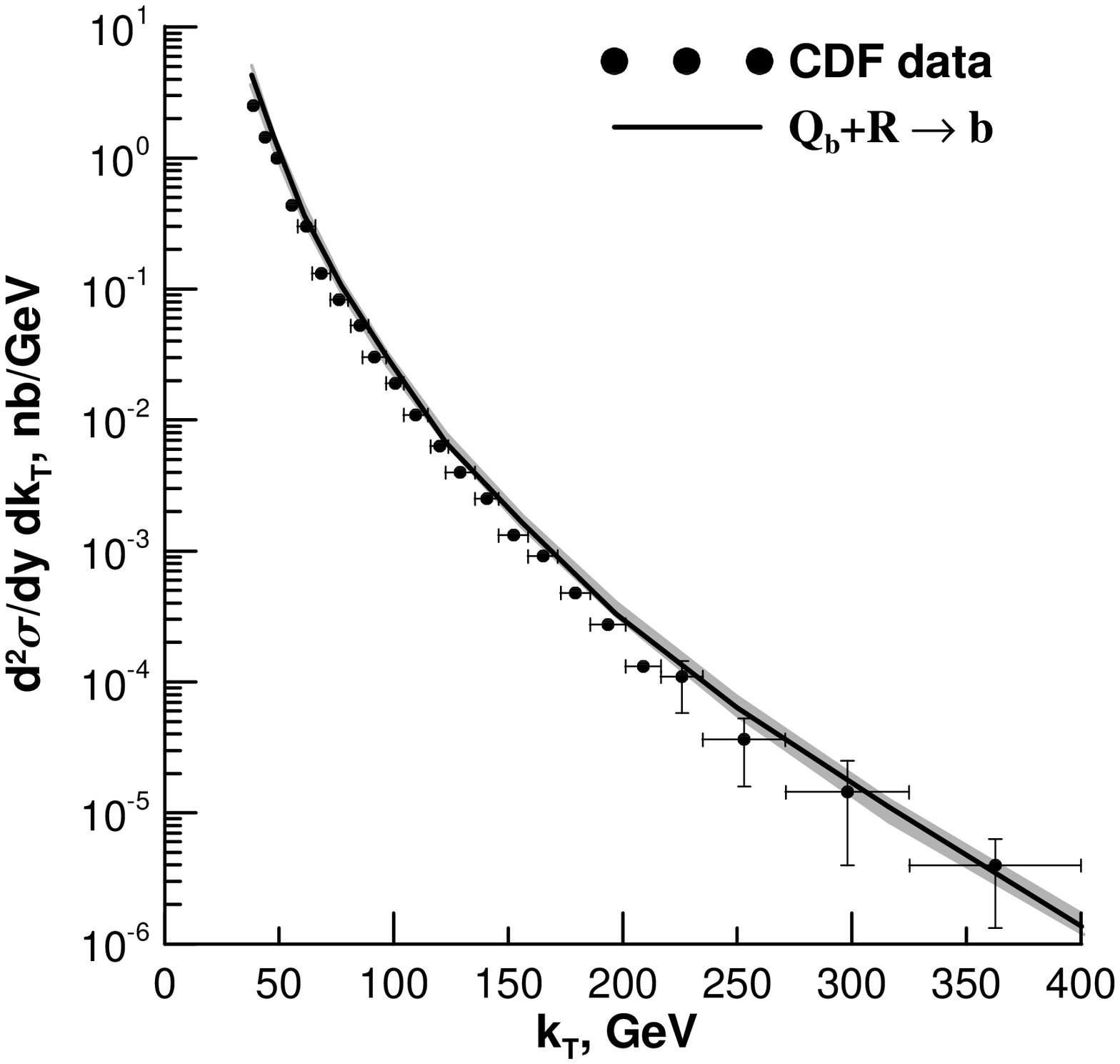}
\end{center}
\caption{\label{fig:3}%
QMRK predictions on inclusive $b$-jet production in comparison
with CDF data \cite{CDF1}.}
\end{figure}

In the Fig. \ref{fig:3} we present the results of our calculation
in comparison with CDF data \cite{CDF1}. One can obtain that they
nicely agree with the CDF data throughout the entire $k_T$ range.

The detailed investigation of single and associated $b$-jet
production in QMRK including these results can be found in our
recent work \cite{KSS2010}.

\section{Inclusive open charm production}
\label{sec:charm} In this section we study $D$-meson production
via charm-quark fragmentation under Tevatron experimental
conditions for the first time in the framework of the QMRK
approach
 complemented with the quark Reggeization hypothesis \cite{KniehlSaleevShipilova}.
The high-energy factorization formulas for these processes are
analogous to ones for inclusive jet production and contain an
extra integral arising from $c\to D$ fragmentation function
$D_{c\to D}$. As for the latter, we adopt the non-perturbative
$D$-meson sets determined in the zero-mass variable-flavor-number
scheme with initial evolution scale $\mu_0=m_c$ \cite{FF} from
fits to OPAL data from CERN LEP1.

The CDF Collaboration \cite{DTevn} measured the $k_T$
distributions of $D^0$, $D^\pm$, $D^*$, and $D_s$ mesons with
rapidity $|y|\le1$ inclusively produced in hadroproduction in
Run~II at the Tevatron, with $\sqrt S=1.96$~TeV. To LO in the QMRK
approach there is only one partonic subprocess
$\mathcal{Q}^{p(\overline p)}_c\mathcal{R}^{\overline{p}(p)}\to
c$, where the top subscript indicates the mother particle. This
subprocess is described via the Reggeized-quark--Reggeized gluon
effective vertex (\ref{amp:QRq}).

\begin{figure}[hb]
\centerline{\includegraphics[width=0.45\textwidth]{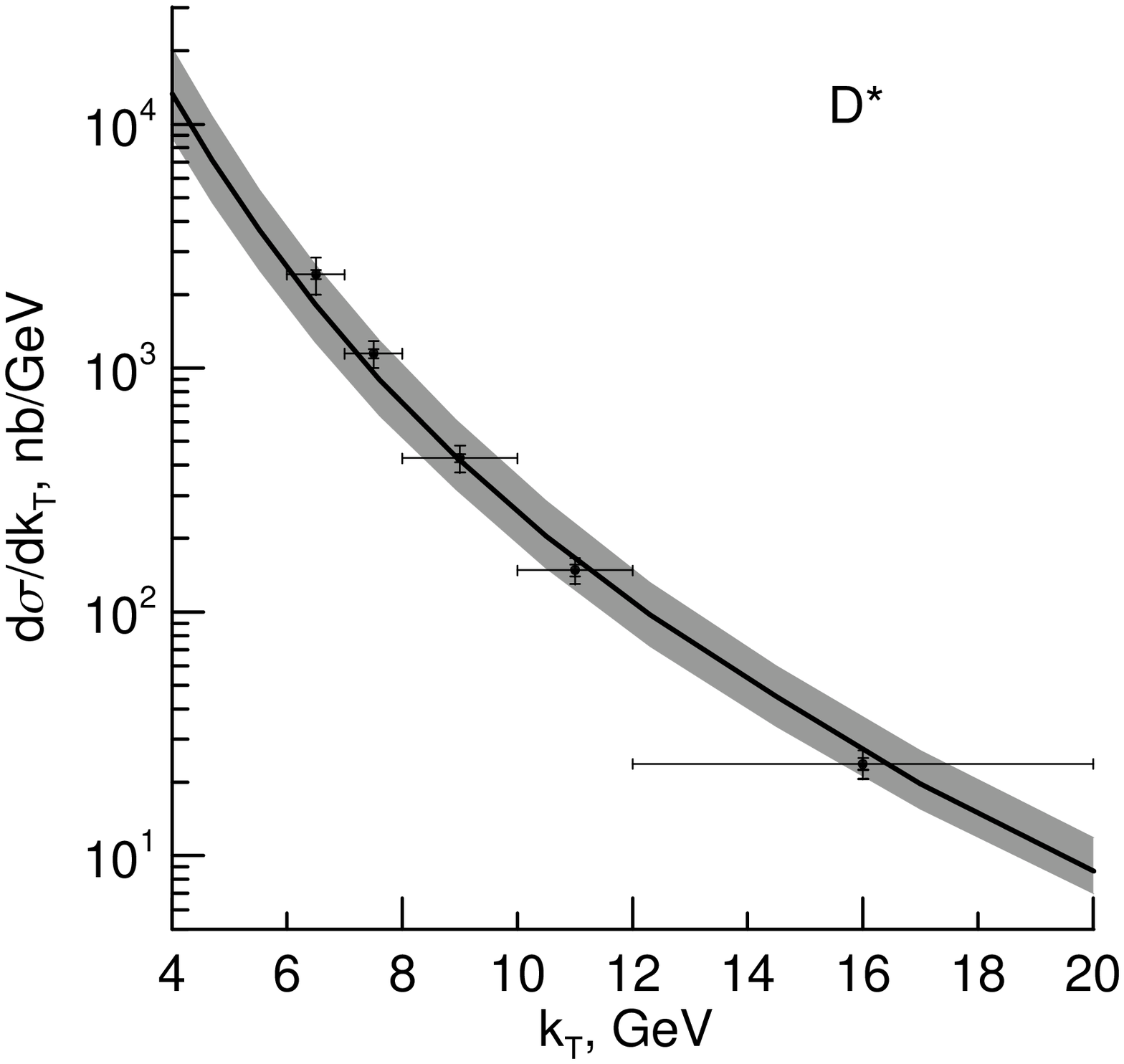}
\includegraphics[width=0.45\textwidth]{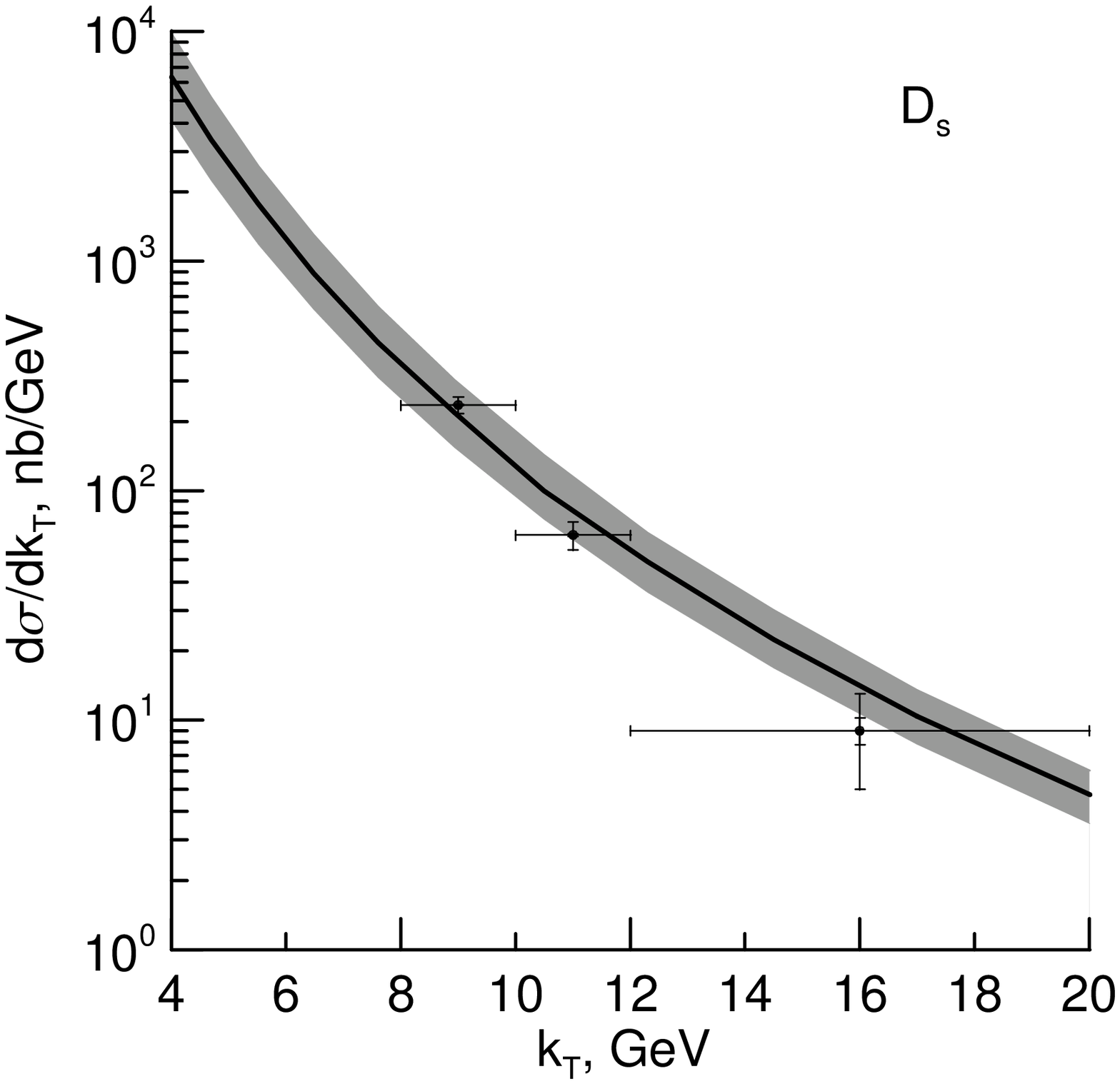}}
\caption{The $k_T$ distributions of inclusive $D^{*}$ (left) and
$D_s$ (right) hadroproduction for $\sqrt S=1.96$~TeV and
$|y|\le1$. The CDF data from Ref.~\cite{DTevn} are compared with
LO predictions from the QMRK approach within the quark
Reggeization hypothesis.}\label{figDTevn}
\end{figure}

In Fig.~\ref{figDTevn}, our results for $D^*$ and $D_s$ mesons are
compared with the CDF data \cite{DTevn}. We find that the
theoretical predictions generally agree rather well with the
experimental data, except perhaps for the slope. In fact, the
predictions exhibit a slight tendency to undershoot the data at
small values of $k_T$ and to overshoot them at large values of
$k_T$ \cite{KniehlSaleevShipilova}. However, we have to bear in
mind that these are just LO predictions, so that there is room for
improvement by including higher orders.

\section{Prompt photon production}
In this part we consider an inclusive production of prompt photons
at Fermilab Tevatron Run~I \cite{QQy1} and Run~II \cite{QQy2}
measured by D0 Collaboration. In the QMRK approach the LO
contribution comes from the Reggeized quark-Reggeized antiquark
annihilation, and we can write the factorization formula as
follows \cite{Saleev2008}:
\begin{eqnarray} d\sigma(p\bar p\to \gamma
X)=\int\frac{dx_1}{x_1}\int
\frac{d^2q_{1T}}{\pi}\int\frac{dx_2}{x_2}\int
\frac{d^2q_{2T}}{\pi} \Phi^p_{q}(x_1,t_1,\mu^2)\Phi^{\bar
p}_{q}(x_2,t_2,\mu^2)
d\hat\sigma(\mathcal{Q}_q\bar{\mathcal{Q}}_q\to \gamma),
\end{eqnarray}
where $q=u,d,s,c,b$, and the charge conjugated subprocesses are
taken into account.
\begin{figure}[ht]
\begin{center}
\includegraphics[width=.7\textwidth, clip=]{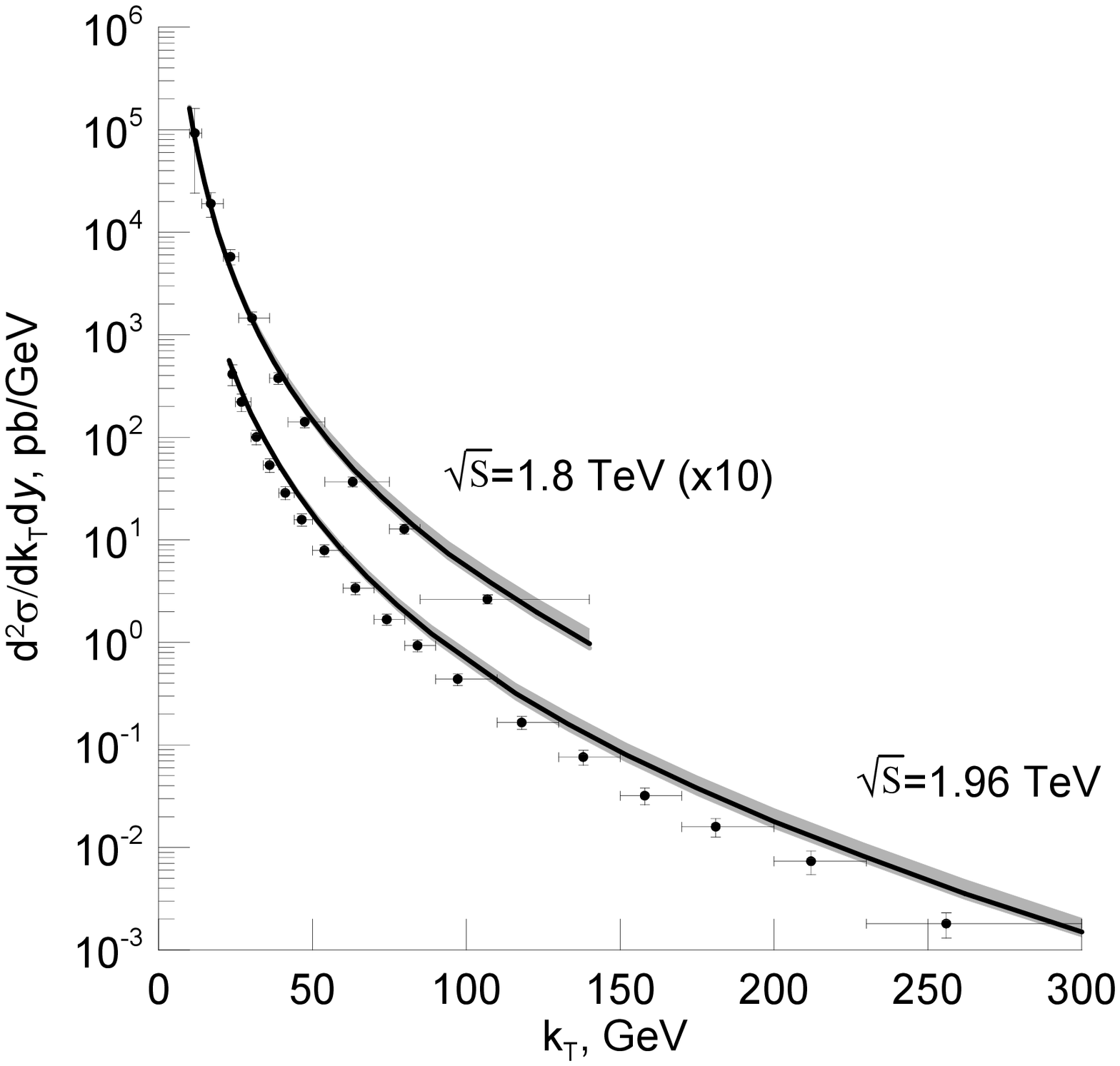}
\end{center}
\caption{\label{fig:4}%
QMRK predictions on prompt photon production in comparison with D0
data \cite{QQy1,QQy2}. }
\end{figure}

In the Fig. \ref{fig:4} we present the results of our calculation
in comparison with the D0 Collaboration data \cite{QQy1,QQy2}. We
see that the agreement between the data and the theoretical
calculation is well again except in the region of $p_T>100$ GeV,
where theoretical results overestimate the experimental data.
Note, that in this region of the photon transverse momentum the
parton longitudinal momentum fractions become non-small and so
long do not satisfy the conditions of particle Reggeization. At
very large $p_T$ one has $x_{1,2}\geq 0.1$ and so far the
collinear parton model should be applied, where the squared
Reggeized amplitude $\overline{|M(\mathcal{Q}\bar{\mathcal{Q}}\to
\gamma)}|^2\to 0$ and the $2\to 2$ parton subprocesses ($qg\to q
\gamma, q\bar q\to g \gamma$, etc.) are needed to take into
account.

\section{Conclusions}
We studied the inclusive hadroproduction of jets, $D$-mesons,
$b$-jets and prompt photons at LO in the QMRK approach, applying
the quark Reggeization hypothesis, under Tevatron experimental
conditions. Despite the great simplicity of our formulas, our
theoretical predictions turned out to describe recent measurements
of transverse momenta cross section distributions of different
final particles at the Tevatron surprisingly well, without any
ad-hoc adjustments of input parameters. By contrast, in the
collinear parton model of QCD, such a degree of agreement can only
be achieved by taking next-to-leading corrections into account and
performing soft-gluon resummation. In conclusion, the QMRK
approach is once again
\cite{Saleev2008,KSS2010,KniehlSaleevShipilova,SaleevDiphoton}
proven to be a powerful tool for the theoretical description of
QCD processes in the high-energy limit.

\section{Acknowledgements}
This work  was supported in part by the Ministry of Science and
Education of Russian Federation, Contract No. P1338. A.~S. is
grateful to the International Center of Fundamental Physics in
Moscow and the Dynastiya Foundation for financial support.


\begin{thebibliography}{99}
\bibitem{QMRK}
  V.~S.~Fadin and L.~N.~Lipatov,
  Nucl.\ Phys.\  {\bf B406}, 259 (1993);
  Nucl.\ Phys.\  {\bf B477}, 767 (1996).

\bibitem{Lipatov95}
  L.~N.~Lipatov,
  Nucl.\ Phys.\  {\bf B452}, 369 (1995).

\bibitem{LipatoVyazovsky}
  L.~N.~Lipatov and M.~I.~Vyazovsky,
  Nucl.\ Phys.\  {\bf B597}, 399 (2001).

  \bibitem{GellMann} M.~Gell-Mann, M.~L.~Goldberger, F.~E.~Low,
E.~Marx, and F.~Zachariasen, Phys.\ Rev.\ \textbf{133}, 145B
(1964).

\bibitem{BFKL}
E.~A.~Kuraev, L.~N.~Lipatov, and V.~S.~Fadin, Sov.\ Phys.\ JETP
\textbf{44}, 443 (1976) [Zh.\ Eksp.\ Teor.\ Fiz.\  \textbf{71},
840 (1976)]; I.~I.~Balitsky and L.~N.~Lipatov, Sov.\ J.\ Nucl.\
Phys.\ \textbf{28}, 822 (1978) [Yad.\ Fiz.\ \textbf{28}, 1597
(1978)].

  \bibitem{QR1old} V.S.\,Fadin, V.E.\,Sherman,
JETP Lett. \textbf{23}, 548 (1976) [Pisma Zh.Eksp.Teor.Fiz.
\textbf{23}, 599 (1976)].
%
\bibitem{QR2old} V.S.\,Fadin,
V.E.\,Sherman, JETP \textbf{45}, 861 (1977) [Zh.Eksp.Teor.Fiz.
\textbf{72}, 1640 (1977)].

\bibitem{Saleev2008}
  V.~A.~Saleev,
  Phys.\ Rev.\  D {\bf 78}, 034033 (2008);
  Phys.\ Rev.\  D {\bf 78}, 114031 (2008).

\bibitem{RRgold} L.N.\,Lipatov, V.S.\,Fadin, JETP Lett. \textbf{49}, 352 (1989)
[Sov. J. Nucl. Phys. \textbf{50}, 712 (1989)].

\bibitem{KMR}
  M.~A.~Kimber, A.~D.~Martin, and M.~G.~Ryskin,
  Phys.\ Rev.\  D {\bf 63}, 114027 (2001).

  \bibitem{Watt} G.~Watt,
URL: {\tt http://www.hep.ucl.ac.uk/\~{}watt/}; G.~Watt,
A.~D.~Martin, and M.~G.~Ryskin, Eur.\ Phys.\ J.\  \textbf{C31}, 73
(2003).

\bibitem{MRST}
  A.~D.~Martin, R.~G.~Roberts, W.~J.~Stirling, and R.~S.~Thorne,
  Phys.\ Lett.\  B {\bf 531}, 216 (2002).

\bibitem{RRgCDF} CDF Collaboration, T.~Aaltonen \emph{et al.},
Phys. Rev. D \textbf{78}, 052006 (2008), Erratum \emph{ibid.} D
\textbf{79}, 119902 (2009).

\bibitem{RRgD0} D0 Collaboration, V.~M.~Abazov \emph{et.al},
Phys. Rev. Lett. \textbf{101}, 062001 (2008).

\bibitem{CDF1} CDF Collaboration, T. Aaltonen {\it et al.},
CDF note 8418 (2006), URL:
\verb$http://www-cdf.fnal.gov/physics/new/qcd/QCD.html$.

\bibitem{KSS2010}
B.~A.~Kniehl, A.~V.~Shipilova, V.~A.~Saleev,
Phys. Rev. D {\bf 81} (2010) 094010.



\bibitem{KniehlSaleevShipilova} B.~A.~Kniehl, A.~V.~Shipilova, and
V.~A.~Saleev, Phys.\ Rev.\ \textbf{D79}, 034007 (2009).

\bibitem{FF}  J.~Binnewies, B.~A.~Kniehl, and G.~Kramer, Phys.\ Rev.\  \textbf{D58}, 014014 (1998);
B.~A.~Kniehl and G.~Kramer, {\it ibid.}\ {\bf 71}, 094013 (2005);
{\bf 74}, 037502 (2006).


  \bibitem{DTevn}
  CDF Collaboration, D.~E.~Acosta {\it et al.},
  Phys.\ Rev.\ Lett.\  {\bf 91}, 241804 (2003).

  \bibitem{QQy1} D0 Collaboration,
B.~Abbott \emph{et al.}, Phys.\ Rev.\ Lett.\  \textbf{84}, 2786
(2000).

\bibitem{QQy2} D0 Collaboration, V.~M.~Abazov \emph{et al.}, Phys.\
Lett.\ \textbf{B639}, 151 (2006).


\bibitem{SaleevDiphoton}
V.A. Saleev, Phys. Rev. D\textbf{80}, 114016 (2009).

\end{thebibliography}
\end{document}